**Semiconducting SiGeSn High-Entropy Alloy: A Density Functional Theory Study**

Duo Wang, Lei Liu, Wenjiang Huang, Houlong L. Zhuang*

School for Engineering of Matter Transport and Energy, Arizona State University, Tempe, AZ 85287, USA

*zhuanghl@asu.edu

**Abstract**

High-entropy alloys (HEAs), which have been intensely studied due to their excellent mechanical properties, generally refer to alloys with multiple equimolar or nearly equimolar elements. According to this definition, Si-Ge-Sn alloys with equal or comparable concentrations of the three Group IV elements belong to the category of HEAs. As a result, the equimolar elements of Si-Ge-Sn alloys likely cause their atomic structures to exhibit the same core effects of metallic HEAs such as lattice distortion. Here we apply density functional theory (DFT) calculations to show that the SiGeSn HEA indeed exhibits a large local distortion effect. Unlike metallic HEAs, our Monte Carlo and DFT calculations show that the SiGeSn HEA exhibits no chemical short-range order due to the similar electronegativity of the constituent elements, thereby increasing the configurational entropy of the SiGeSn HEA. Hybrid density functional calculations show that the SiGeSn HEA remains semiconducting with a band gap of 0.38 eV, promising for economical and compatible mid-infrared optoelectronics applications. We then study the energetics of neutral single Si, Ge, and Sn vacancies and (expectedly) find wide distributions of vacancy formation energies, similar to those found in metallic HEAs. However, we also find anomalously small lower bounds (e.g., 0.04 eV for a Si vacancy) in the energy distributions, which arise from the bond reformation near the vacancy. Such small vacancy formation energies and their associated bond reformations retain the semiconducting behavior of the SiGeSn HEA, which may be a signature feature of a semiconducting HEA that differentiates from metallic HEAs.

**Introduction**

Conventional alloy designs often start with selecting one element as the principal constituent and adding other elements to optimize the properties. After centuries of intense efforts on developing these alloys, obtaining targeted properties for traditional alloys is approaching its limit [1]. HEAs represent a new type of alloys that can potentially break the limit due to the presence of a variety of attractive properties currently absent from conventional alloys, making them the center of immense attention [2]. For instance, it is well known that an optimal engineering alloy requires a trade-off between toughness and strength, as these two properties favor and disfavor the movement of dislocations, respectively. Li et al. recently showed that novel FeMnNiCoCr HEAs can overcome this trade-off [3]. Other desirable properties of HEAs include antioxidant capacity [4], high temperature strength [5], high corrosion resistance [6, 7], etc.

Two main definitions of HEAs are commonly used in the literature. The first one, proposed in 2004 by Yeh et al., is based on composition [8], which states that HEAs refer to the alloys containing at least five principal elements, each of which has an atomic concentration between 5% and 35%, rather than unnecessary constraints, i.e., equimolar or near-equimolar concentration. Furthermore, similar to conventional alloys, HEAs may also contain minor elements (i.e., atomic concentration less than 5%) in order to tune the properties of base systems, which further expands the number of possible HEAs [9]. The second definition focuses on the magnitude of the configurational entropy. Because configurational entropy is often dominant in the total entropy of a system in comparison with the other entropies, i.e., vibrational, magnetic, and electronic entropies [10], the total entropy is approximated by the configurational entropy to avoid expensive calculations of the other entropies [11]. According to Boltzmann's entropy formula, the ideal configurational entropy of mixing per mole $\Delta S_{mix}$ for an ideal random $N$-component solid solution can be written as,

$$\Delta S_{mix} = -R\sum_{i=1}^{N} c_i \ln c_i, \qquad (1)$$

where $R$ (8.31 J/K·mol) is the gas constant, $c_i$ refers to the atomic concentration of the $i^{th}$ element, and $N$ is the total number of elements. If $N$ is fixed, the maximum $\Delta S_{mix}$ is achieved when the atomic concentration for all the elements is the same. The second definition therefore implies that HEAs favor equimolar composition. Moreover, this definition further separates HEAs into low ($\Delta S_{mix} < 0.69\ R$), medium ($0.69\ R < \Delta S_{mix} < 1.61\ R$), and high ($\Delta S_{mix} > 1.61\ R$) entropy alloys [12]. Yeh recently suggested that the boundary entropies ($0.69\ R$ and $1.61\ R$) are replaced by more reasonable $1.0\ R$ and $1.5\ R$, respectively [13].

HEAs generally exhibit four phases: solid solution, intermetallic compound (i.e., a compound with a specific stoichiometry), mixed solid solution and intermetallic compound, and bulk metallic glasses. HEAs with a solid solution phase are often preferred, as most desirable properties are associated with this phase [14]. According to the entropy-based definition of HEAs, the stability of a solid solution phase can be enhanced by increasing the number of elements to maximize the configurational entropy [15, 16]. But the probability of at least one pair of elements forming the intermetallic phase is directly proportional to the number of elements, leading to the competition between the solid solution and intermetallic phases [17, 18]. It is therefore a daunting task to design new HEAs, i.e., to predetermine the phase given a combination of elements and the concentrations that form a HEA.

HEAs also show four core effects, which are the high configurational entropy effect, the sluggish diffusion effect, the lattice distortion effect, and the "cocktail" effect. These effects describe HEAs from the aspects of thermodynamics, kinetics, structures, and properties, respectively [12, 19]. Based on the second law of thermodynamics, the high configurational entropy effect lowers the Gibbs free energy by compensating for the enthalpic change in the system, leading to possible formations of stable phases. Depending on the competition between enthalpy and entropy, both disordered solid solution phases and ordered intermetallic compounds can form a HEA [20]. The high entropic effect favors a disordered solid solution phase. The sluggish diffusion effect kinetically lowers the rate of atomic diffusion, and thus reduces the overall phase transformation rate in an HEA, in contrast to conventional alloys. Because atoms in HEAs are usually bonded with the atoms of other elements, most of the atoms experience different diffusion paths and have different diffusion barriers [19, 21]. Furthermore, forming an HEA by introducing multiple elements with different atomic sizes is associated with the lattice distortion effect. This originates from lattice strain and stress, as different elements have their own atomic radii, bonding energies, and structural preferences. The lattice distortion effect affects properties of an HEA such as hardness, electrical and thermal conductivity [11, 12]. The "cocktail" effect of HEAs refers to the enhancement of material properties due to the presence of multiple principal components [13]. This effect emphasizes not only the individual elemental advantage but the synergetic results from the interactions among the atoms of multiple elements.

Significant research efforts have been devoted to studying the mechanical properties of HEAs for a variety of engineering applications. Even though HEAs vary widely from the constituent elements and their compositions, many of them commonly show useful mechanical properties such as high hardness values [22-25], yielding stresses [26-28], fatigue resistance [29-31], and irradiation resistance [32, 33].

In contrast to the work on the mechanical properties of HEAs, there are much fewer studies on their functional properties such as magnetic and semiconducting properties [34]. Recently, several studies have started to discover these functional properties in HEAs. For example, by introducing metallic elements including ferromagnetic Fe, Co, and Ni, the resulting HEAs display paramagnetic [35] or even superparamagnetic properties [36]. Generally, introducing non-magnetic elements into HEAs as additional principal components also impacts the magnetism of the original HEAs. The change of magnetism in HEAs depends on the included elements, which cause a structural change and the formation of a solid solution phase. Recent studies also indicate that the saturation magnetization at room temperature and the Curie temperature of HEAs are both tunable by controlling the concentrations of the principal elements [37]. As an example of the semiconducting properties of HEAs, recent studies found that upon successively adding alloying elements Ge, Pb, and Mn to the SnTe binary alloy to form HEAs, the valence bands and the band gaps in the HEAs are modified as a result of the cocktail effect [38]. In particular, the ternary Sn-Ge-Te HEA showed a drastic reduction in the band gap without significantly modifying the original band structure. On the contrary, the Sn-Ge-Pb-Mn-Te HEA not only had a widened band gap, but also more flattened valence bands in the band structure than those in the binary SnTe alloy [38].

In this work, we aim to extend the study of semiconducting properties of HEAs. We choose HEAs consisting of group IV elements (Si, Ge, and Sn) because of the important roles played by this group of alloys in the optoelectronic devices. According to the definition, it is more accurate to classify the SiGeSn alloy as a medium-entropy alloy (MEA). Nevertheless, MEAs share many common properties with HEAs, so we call the SiGeSn HEA throughout the current work. Separate from the context of HEAs, this group of (binary or ternary) alloys have continuously attracted massive attention over the past several decades. For example, $Ge_{1-x}Sn_x$ alloys display tunable band gaps if the concentration of Ge or Sn is controlled. It is reported that an Sn content up to 8% not only lowers the band gap, but also changes the indirect-band-gap Ge to a direct-band-gap semiconducting alloy [39]. A similar effect is observed in Si-Sn alloys [40]. Inspired by these binary alloys, research on the $Si_xGe_{1-x-y}Sn_y$ ternary alloys also reveals the Sn-content-dependent band gaps as well as the transition between direct and indirect band gaps [41].

Experimentally, it has been challenging to obtain Si-Ge-Sn alloys especially with high Sn content. The difficulty of alloying Si-Ge alloys with Sn can be understood from the three binary (Si-Ge, Si-Sn, and Ge-Sn) phase diagrams. First, the Si-Ge phase diagram shows that any composition of Ge is completely soluble in Si, forming Si-Ge solid solution alloys [42]. Due to this solubility, a number of Si-Ge alloys have been experimentally developed, resulting in a wide range of applications such as in near-infrared devices [43]. By contrast, as shown in the Si-Sn and Ge-Sn phase diagrams [44, 45], the solution limit at room temperature is below 1%. The striking differences in the solubility of Ge and Sn in Si can be understood from the Hume-Rothery rules [46], which are commonly used to predict whether a binary alloy prefers to exhibit a solid solution phase or an intermetallic compound. Because of the limited number of the experimental data, how Hume-Rothery rules can be applied to ternary alloys remains unclear. Nevertheless, we can apply the Hume-Rothery rules to gain an intuitive understanding of Si-Ge-Sn ternary alloys. According to Hume-Rothery, four conditions need to be satisfied to form binary alloys with a solid solution phase: (i) mismatch in atomic radii should not exceed 15%, i.e., size effect; (ii) there must be a similarity between the crystal structures of solute and solvent; (iii) if the solute and solvent have the same valency, complete solubility occurs, i.e., valency effect; (iv) the electronegativity of solute and solvent should be similar. Si and Ge satisfy all four of these conditions: possessing the same cubic diamond structure, similar atomic radii (1.153 Å for Si and 1.24 Å for Ge [47]), same valency (+4), and similar electronegativity (1.90 for Si and 2.01 for Ge at the Pauling scale [47]). Thus, complete solubility exists in Si-Ge binary alloys. For Si-Sn, condition (i) is not satisfied, due to the atomic radius of Sn (1.62 Å [47]) being too large. Condition (ii) is satisfied only at low temperatures, where Sn crystalizes as the same crystal structure called the $\alpha$-Sn phase, while at higher temperatures a phase transitions occurs and the structure is transformed to a tetragonal structure, i.e., the $\beta$-Sn phase. Conditions (iii) and (iv) are both met; the electronegativity of Sn is 1.96 at the Pauling scale [47]. The net effect of these four conditions for Si-Sn is that Si and Sn do not form a solid solution phase. The same net effect applies to the Ge-Sn binary system.

Intensive experimental developments have shown how a higher content of Sn can now be included in Si-Ge-Sn alloys [48]. Several innovative methods have been developed to fabricate high-Sn content Si-Ge-Sn alloys, including molecular beam epitaxial (MBE) [49] and chemical vapor deposition method (CVD) [50]. However, the limit of Sn content in these Si-Ge-Sn alloys is unknown. And if such a limit exists, what is the electronic structure of the Si-Ge-Sn alloy? In addition, there is no answer to whether the phase transition to the (metallic) $\beta$-Sn phase occurs in high-Sn content Si-Ge-Sn alloys.

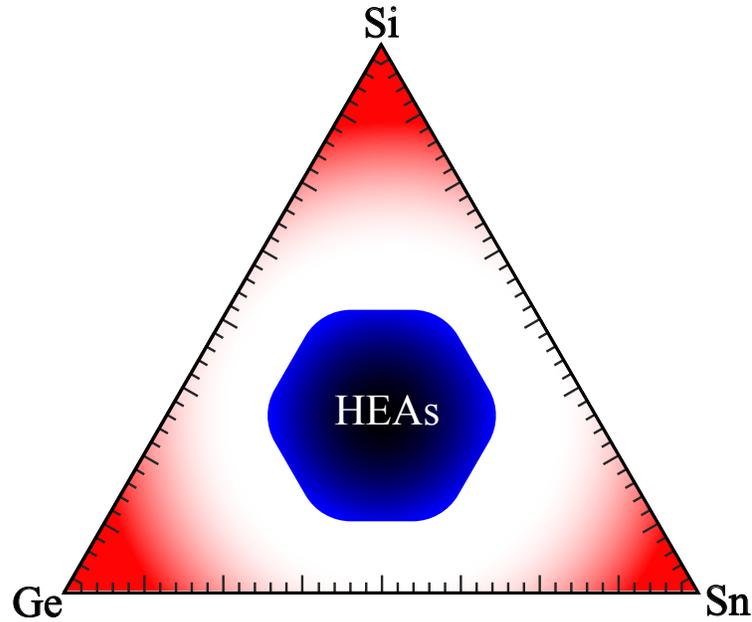

**Figure 1.** Schematic of a Si-Ge-Sn ternary phase diagram. Red shaded areas correspond to the composition spaces of conventional Si-Ge-Sn alloys; blue shaded areas correspond to the Si-Ge-Sn high-entropy alloys (HEAs).

The previous studies on Si-Ge-Sn alloys are constrained to a small area of the composition space. In other words, only three corner regions of the Si-Ge-Sn composition space have been exploited (see Figure 1). In these corner regions, Si-Ge-Sn alloys are referred to as conventional alloys with a dominant element such as Ge, which has a content higher than 70%, and the other two elements in the conventional alloy regarded as doping elements to tune the alloys' properties. Indeed, a significant amount of research has been performed that focuses on the mechanical [51, 52], optical [53], and electrical properties [52, 54, 55] of the alloys at the corners of the Si-Ge-Sn ternary phase diagram, with the Si content between 10% to 20% and the Sn content no more than 10%. On the contrary, little research has been carried out for the Si-Ge-Sn alloys in the middle region of the ternary phase diagram. According to the definition of HEAs mentioned above, the Si-Ge-Sn alloys in the middle region of the composition space should be HEAs. Interestingly, this terminology has never been used for Si-Ge-Sn alloys in the literature.

Similar to studying metal alloys in the center region of the composition space, there are enormous opportunities and challenges to study Si-Ge-Sn alloys in this region from both experiment and theory points of view. On the experimental side, it is worthwhile to fabricate the Si-Ge-Sn HEAs, measure their electronic structures, and to see if they are different from the conventional Si-Ge-Sn alloys. However, the setup of the CVD method may need to be redesigned to accompany the near-equimolar or equimolar content of the three elements. Theoretically, it is a great example of using *ab initio* density functional theory (DFT) [56, 57] calculations to predict properties before an Si-Ge-Sn HEA is successfully fabricated. From the structural perspective, Si-Ge-Sn HEAs are intrinsically associated with the four core effect, which affect the electronic structures, potentially leading to development of structures offering electromagnetic spectrum dominance. Meanwhile, because of the solid solution phase assumed in the HEAs, we need to employ reasonable supercell models to simulate them.

In this work, we first compute the electronic structure of the SiGeSn HEA to determine whether it is metallic or semiconducting. We then examine if the HEA exhibits a chemical short-range order (CSRO) which is a tendency for atomic clustering (e.g., one Si atom prefers Si/Ge neighbors over Sn atoms). CSRO has been reported in numerous HEAs and dominates several of their properties [58, 59]. Furthermore, because point defects are unavoidable in any material due to thermal vibrations, we study the most basic point defect in the SiGeSn HEA, i.e., single vacancies. Studying single vacancies allows an understanding at an atomic level which can provide guidance to future studies of doping the SiGeSn HEAs.

**Methods**

We use the Vienna *ab-initio* simulation package (VASP; version 5.4.4) for all the DFT calculations and the Perdew-Burke-Ernzerhof (PBE) functional to describe the exchange-correlation interactions [60]. We also use the standard Si, Ge, and Sn potential datasets based on the PBE functional and the projector-augmented wave (PAW) method [61, 62] to describe the electron-nuclei interactions. Among the potentials, the $3s^2$ and $3p^2$ electrons of Si atoms, the $4s^2$ and $4p^2$ electrons of Ge atoms, and the $5s^2$ and $5p^2$ electrons of Sn atoms are regarded as valence states. We optimize the supercells using a $2 \times 2 \times 2$ Monkhorst-Pack [63] $k$-point grid and a cut-off energy of 400 eV for the plane wave basis sets. The force convergence criterion is set to 0.01 eV/Å.

We use a special quasi random structure (SQS) as a starting point to simulate the crystal structure of the SiGeSn HEA. The SQS method was developed nearly three decades ago, and has been used in many applications for modeling conventional semiconducting alloys (e.g., $Al_{1-x}Ga_xAs$) with small supercells that can be dealt with using standard DFT programs [64]. The goal of this method is to minimize the difference between the correlation functions in a small supercell and those in an alloy with a truly random structure. Recently, the SQS method has been widely used to simulate the structures of metallic HEAs with a solid solution phase [25]. In addition, an SQS structure also serves as an initial structure followed by a combination of MC and DFT simulations to examine the occurrence of CSRO, which has been reported to occur in typical metallic HEAs such as Cr-Co-Ni [58, 59] and affect mechanical properties such as stacking fault and point-defect energies. We focus on the SiGeSn HEA with the highest Sn content (i.e., where the ratio of the three elements is 1:1:1).

Based on an MC procedure, we use the mcsqs module [65] implemented in the Alloy Theoretical Automated Toolkit (ATAT) package [66] to generate an SQS structure for the SiGeSn HEA with 216 atoms (corresponding to a $3 \times 3 \times 3$ supercell of the 8-atom unit cell of Si, Ge, or $\alpha$-Sn). The cutoff distance for computing the correlation functions is set to within the second nearest-neighbor (NN) bond length, as a larger cutoff distance leads to an unconverged calculation. The SQS structure is then fully relaxed and the shape of the supercell is slightly off perfectly cubic. Figure 2 shows a zoomed-in view of the optimized SQS structure. As clearly shown, the four NN atoms of each atom often belong to different elements, confirming the randomness of the atomic positions in the SQS structure. These results represent the SiGeSn HEA with a random solid solution phase.

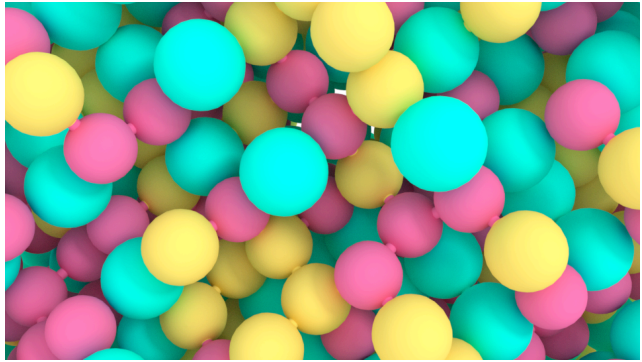

**Figure 2.** Zoomed-in view of the SQS structure of a SiGeSn high-entropy alloy. Si, Ge, and Sn atoms are represented by pink, yellow, and cyan spheres, respectively.

**Results and Discussion**

We first benchmark the above settings by calculating the optimized lattice constants of diamond cubic Si, Ge, and α-Sn as 5.47, 5.78, and 6.65 Å, respectively, which are slightly higher than the corresponding experimental lattice constants of 5.43, 5.66, 6.49 Å respectively [67]. The difference is expected from using the PBE functional that generally overestimates the lattice constants [68]. We also calculate the cohesive energies of isolated Si, Ge, and Sn atoms place in a vacuum box with unequal side lengths of 22.0, 23.0, and 24.0 Å (to break the symmetry, spin-polarized calculations are invoked). For all the isolated atoms, we obtain a spin magnetic moment of 2 $\mu_B$, which match the experimental atomic spectra label $^3P_0$. The cohesive energy, i.e., the energy required to break the Si-Si, Ge-Ge, and Sn-Sn bond into isolated atoms is calculated as 4.60, 3.73, and 3.18 eV, respectively, which agrees with the literature values of 4.63, 3.85, and 3.14 eV, respectively [67]. The cohesive energies show that the bond strengths follow the order: Si-Si > Ge-Ge > Sn-Sn bond. Furthermore, we calculate the PBE band gap of diamond cubic structure Si as 0.61 eV (consistent with previously reported 0.75 eV [69]), whereas Ge and $\alpha$-Sn are predicted to be metal and a semimetal with no gap at the PBE level of theory, consistent again with previous DFT calculations [69, 70]. These benchmark calculations validate our DFT simulation parameters to be used throughout the current work.

To further show that the SQS structure is a reasonable model to simulate the SiGeSn HEA, we begin with the fully optimized SQS structure and perform a combination of DFT and MC calculations. For the DFT calculations, we calculate static energies—similar to those reported in Refs. [58, 59]—to examine whether swapping two atoms of different elements lowers the energy of the SQS structure. This enables us to simultaneously examine whether the SCO occurs in Si-Ge-Sn HEA. We consider three cases of exchanging the atoms: (i) Case 1, by swapping Si with Ge and Sn atoms with an equal probability, (ii) Case 2, by swapping Si with Ge atoms, and (iii) Case 3, by swapping Si with Sn atoms. In the MC simulations, if the exchange of two atoms leads to a lower energy, the atomic exchange is accepted. If not, it is accepted with the probability calculated by comparing the Boltzmann factor ($e^{-\Delta E/k_\mathrm{B}T}$; $\Delta E$: energy change; $k_\mathrm{B}$: Boltzmann constant) at the temperature ($T$) of 300 K with a random number between 0 and 1.

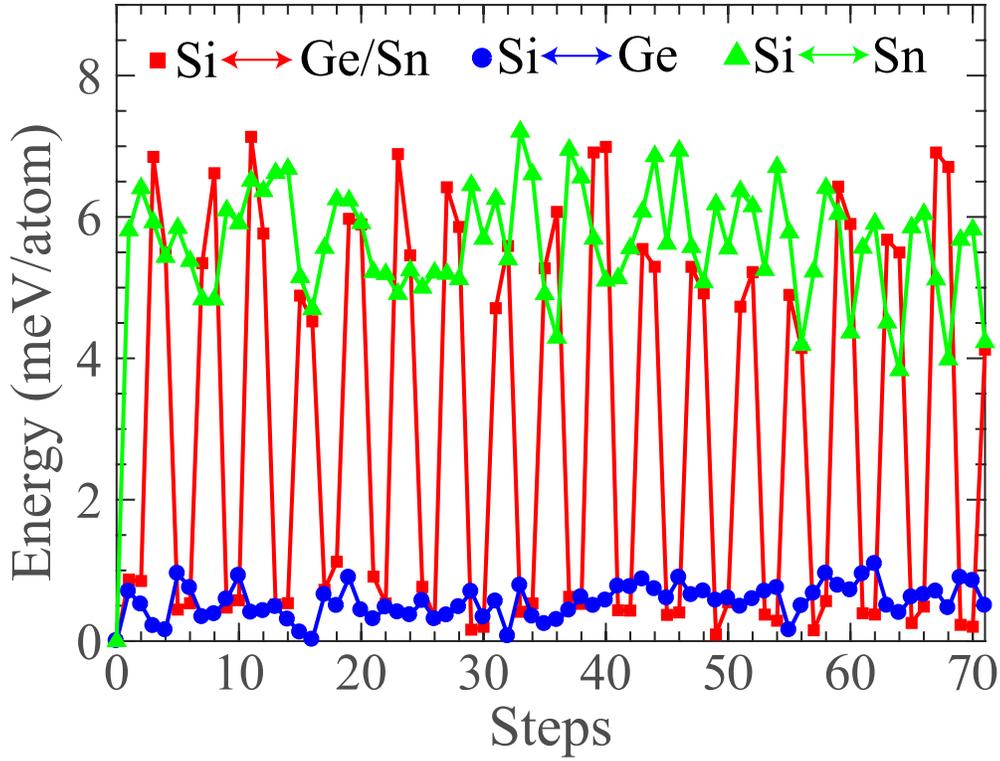

**Figure 3.** Energy change of a special quasi-random structure of the SiGeSn high-entropy alloy as a function of Monte Carlo simulation steps.

Figure 3 displays three curves (labeled by Si↔Ge/Sn, Si↔Ge, and Si↔Sn, respectively) showing the energy changes with reference the SQS structure (set to zero) as the MC calculations progress. We observe that all the energy changes are positive, indicating that the initial SQS structure is the most stable structure. Furthermore, the energy changes by swapping Si and Ge atoms (Case 2) are seen to be much smaller than those by swapping Si and Sn atoms Case 3). The energy changes in Case 2 are close to zero, showing that it is almost energy-free to exchange Si and Ge atoms. This phenomenon is consistent with the complete solubility of Si and Ge resulting from their same valency and relatively small atomic size mismatch. The Si↔Ge/Sn curve (Case 1) shows that every two data points of low energy changes (similar to those in the Si↔Ge curve) are followed by two data points of high energy changes (similar to those in the Si↔Sn curve). This trend of alternating high and low energy changes is due to the fact that in the MC simulation for Case 1, each Si atom is first exchanged with two random Ge atoms, and then exchanged again with two random Sn atoms. As mentioned before, the former exchange results in little difference in the total energy changes, while the latter significantly increases the total energies by approximately 6 meV/atom.

The combined DFT and MC calculations reveal that no CSRO is observed in the SiGeSn HEA, in contrast to the metallic CrCoNi HEA exhibiting CSRO [58, 59]. We suggest that the absence of CSRO in the SiGeSn HEA is because of the similarity in the electronegativity of the three elements. As a result, there is no stable (with negative formation energies) binary or ternary intermetallic compound formed from bulk Si, Ge, or $\alpha$-Sn. By contrast, according to the Materials Project [71], stable binary intermetallic compounds in the CrCoNi HEA such as $CrCo_3$, $CrNi_2$, and $Co_3Ni$ exist with the formation energies of -0.006, -0.021, and -0.022 eV/atom, respectively, which may enhance the tendency of CSRO in the metallic CrCoNi HEA. It is now well accepted that CSRO significantly reduces the configurational entropy [72, 73]. In this regard, the semiconducting SiGeSn HEA should have a higher configurational entropy than the metallic CrCoNi HEA.

Having confirmed the SQS structure as a reasonable one for simulating the SiGeSn HEA, we study the electronic structure of the HEA. Figure 4 shows the density of states of the SiGeSn HEA calculated with the PBE and HSE06 [74] hybrid density functionals. As can be seen, the calculations based on the PBE functional predict the SiGeSn HEA to be metallic, whereas the calculations using the more accurate HSE06 functional show that the SiGeSn HEA is actually a semiconductor with a small band gap of 0.38 eV. Many optoelectronic applications such as nigh vision [75], thermal imaging [76], and biomedical sensing [77] requires band gaps in the mid-infrared region (2.5-10 μm [78] i.e., 0.12-0.50 eV). Therefore, the SiGeSn HEA may be useful for these applications and has the advantage of over the commonly used mid-infrared (group III-V and II-VI) materials in costs [78] and in compatibility with the complementary metal-oxide-semiconductor (CMOS) technology [48].

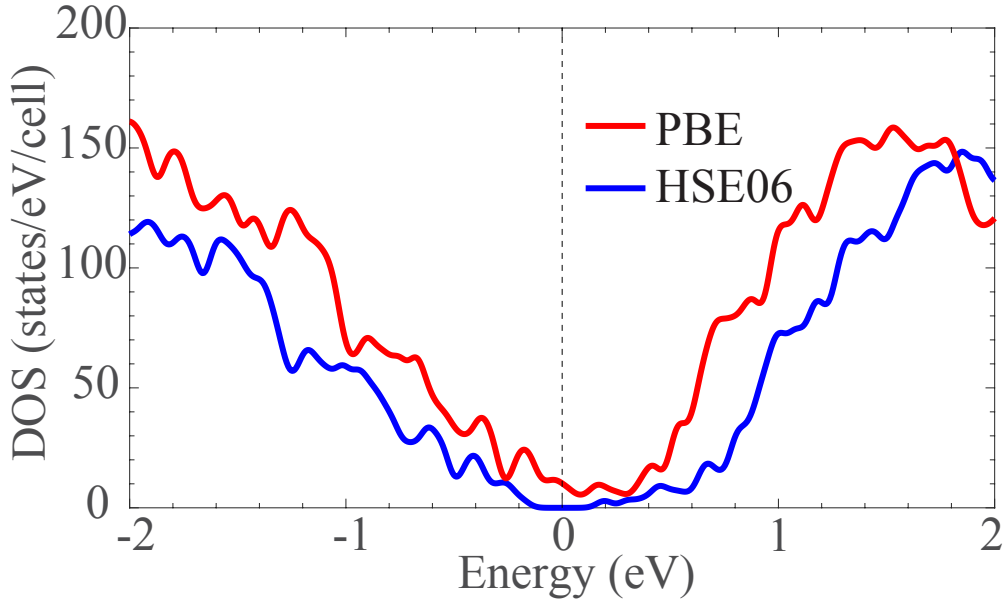

**Figure 4.** Density of states (DOS) of the SiGeSn high-entropy alloy calculated with the PBE and HSE06 functionals.

The lattice distortion effect is ubiquitous in HEAs, so it has become one of the four core effects of HEAs. We here quantify this effect for the SiGeSn HEA. We follow the same method used by Tong et al. for the FeCoNiCr and FeCoNiCrPd HEAs to provide a statistical description of the local distortion effect in SiGeSn [79]. We calculate the distribution of the deviation $\Delta d$ of the NN bond lengths with reference to the average NN bond length $d_{\text{avg}}$ for the optimized SQS structure. Figure 5 displays the distribution $\Delta d/d_{\text{avg}}$ and its Gaussian fit. The standard deviation of the SiGeSn alloy is determined to be 4.46%, which is larger than those (1.04% and 3.37%, respectively) of the FeCoNiCr and FeCoNiCrPd HEAs. We also apply the hard sphere-model proposed by Zhang et al. to determine the size mismatch $\delta$ calculated using the following equation [80],

$$\delta = 100 \times \sqrt{\sum_{i=1}^{N} c_i \left(1 - r_i / \sum_{j=1}^{N} c_j r_j \right)^2}, \qquad (2)$$

where $N$ is the total number of elements, $c_{i/j}$ are the concentrations of the element in an HEA—$c_{i/j}$ is equal to 1/3 for the SiGeSn HEA, and $r_{i/j}$ represent the atomic radii of Si, Ge, and Sn atoms. $\delta$ in Eq. (2) is structure independent, i.e. regardless of the atomic arrangement in an alloy being ordered or not. The resulting $\delta$ based on Eq. (2) is 15.16%, consistent with the large standard deviation of the $\Delta d/d_{avg}$ data; both metrics ($\Delta d/d_{avg}$ and $\delta$) reflect the severe local lattice distortion effect.

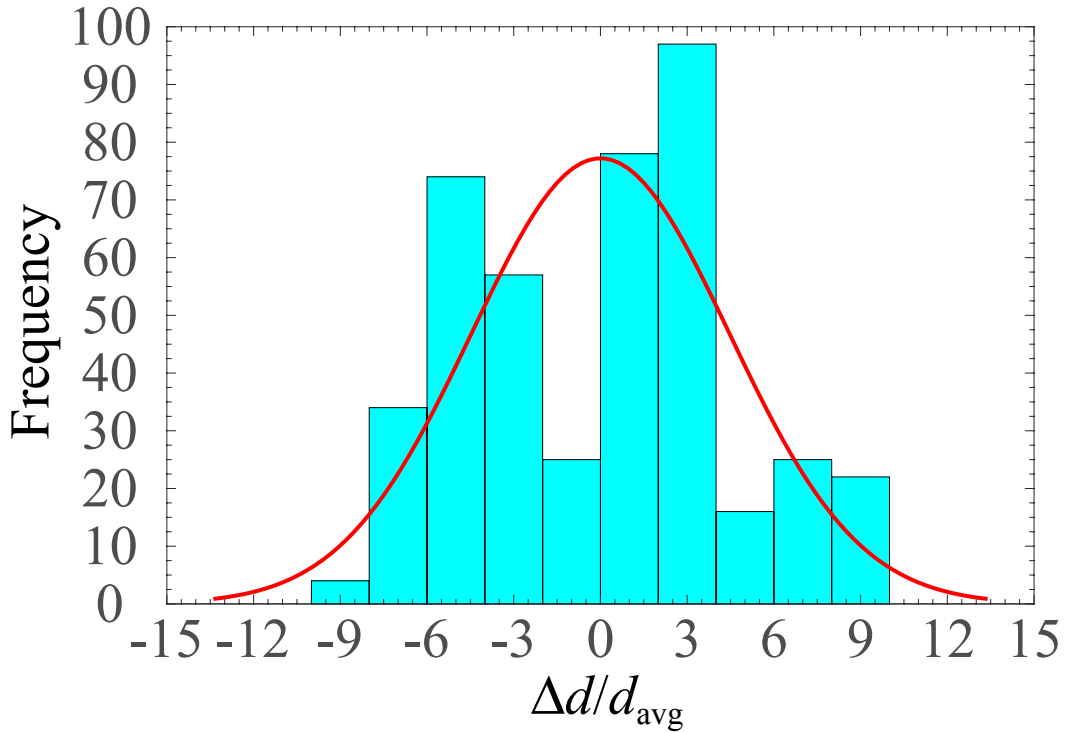

**Figure 5.** Frequencies in the deviation $\Delta d$ of nearest-neighbor (NN) bond lengths from the average NN bond length in the optimized structure of a SiGeSn HEA.

With the optimized SQS structure, we study the vacancy formation energy of a single vacancy and its dependence on the elements and locations of the vacancy. Unlike in metals, vacancies in semiconductors (e.g., the SiGeSn HEA according to our HSE06 calculations) can exhibit various charge states [81]. Additional terms including the valence band maximum and band gap of a supercell without a defect, and potential alignment between the charged defect and systems should be included as a correction for the energy difference between a supercell with charged defect and a supercell without the defect [82]. While the valence band maximum and band gap can be obtained from HSE06 calculations, the band alignment term depends on the dielectric constant of the SiGeSn HEA structure. Using the PBE functional to calculate this parameter is problematic, as the SiGeSn HEA is metallic, exhibiting an infinite dielectric constant. On the other hand, although the HSE06 functional leads to the conclusion that the SiGeSn HEA is semiconducting (and therefore a finite value of dielectric constant), obtaining a converged dielectric constant requires a very dense grid of $k$ points and advanced theories like density functional perturbation theory [83]. Due to these technical issues associated with time-intensive calculations, we therefore consider only the vacancies without charges in this work. We defer the study of charged vacancy defects, which an important issue, to the future work.

Focusing on the neutral vacancy, the single vacancy formation energy $\Delta E$, describing the energy change caused by removing one atom from the bulk and placing it in a reservoir, is defined as [84],

$$\Delta E = E_{\text{vac}} + \mu - E_{\text{bulk}}, \tag{3}$$

where $E_{\text{vac}}$ is the energy of the supercell with a single vacancy, $\mu$ is the chemical potential of an atom in the reservoir, and $E_{\text{bulk}}$ is the energy of the supercell without a vacancy. Because there are no known intermetallic compounds formed among Si, Ge, and $\alpha$-Sn, we assume the reservoir to be bulk Si, Ge, and $\alpha$-Sn for the corresponding vacancies. By the definition shown in Eq. (3), we can compare the vacancy formation energies in the SiGeSn HEA and in bulk Si, Ge, or $\alpha$-Sn on the same footing. Since the chemical potential is the same as the atomic energy in pure bulk, $\Delta E$ for bulk Si, Ge, or $\alpha$-Sn in Eq. 3 is reduced to

$$\Delta E = E_{\text{vac}} - \frac{n-1}{n} E_{\text{bulk}}. \tag{4}$$

Here, $n$ = 216 is the number of atoms in a supercell that is sufficiently large to compute the formation energy of an isolated, neutral vacancy. For all the supercells with a single vacancy, we completely optimize the lattice constants and atomic coordinates with the PBE functional. Table 1 shows the calculated single vacancy formation energies of bulk Si, Ge, and $\alpha$-Sn, which are consistent with the literature. The order of the three vacancy formation energies also agrees well with the order of the cohesive energies, i.e., higher cohesive energy indicates higher vacancy formation energy.

Because the bonding environment of each atom in the SiGeSn HEA is different, the corresponding vacancy formation energy is likely to be distinct. To simulate the three types (Si, Ge, and Sn) of vacancies in the SiGeSn HEA, we create a single vacancy by consecutively removing one of the 216 atoms and fully optimize the resulting defected 215-atom supercell structure, followed by computing the corresponding $\Delta E$. Figure 6 shows the three ranges of $\Delta E$ in the SiGeSn HEA and the ranges are summarized in Table 1. We can see from Figure 6 that the distribution of formation energies for each type of vacancy is nearly continuous. Such wide ranges of vacancy formation energies are expected to some extent and have also been observed in other HEAs. For example, Chen et al. find that the range of the vacancy formation energies of Fe in FeCoCrNi HEAs is from 0.72 to 2.89 eV [85].

**Table 1.** Vacancy formation energies (in eV) of Si, Ge, and Sn in their own bulk crystals and the SiGeSn HEA calculated in this work. For comparison, the experimental and theoretical data of the vacancy formation energies of Si, Ge, and $\alpha$-Sn in the literature are also shown. For the vacancy formation energies of Si, Ge, or Sn in the SiGeSn HEA, we report a range instead of a single value due to the fact that the energy cost of removing an atom from the HEA depends on the location of the atom.

| $V_{Si}^{Si}$ | $V_{Ge}^{Ge}$ | $V_{Sn}^{Sn}$ | $V_{Si}^{HEA}$ | $V_{Ge}^{HEA}$ | $V_{Sn}^{HEA}$ |
|---|---|---|---|---|---|
| 3.83 | 2.08 | 1.33 | 0.04-2.58 | 0.20-2.37 | 0.39-2.55 |

| | |
|---|---|
| 4.0[a] | 2.04-2.62[b], |
| | 2.35[c] |

[a]Ref. [86]. Experimental work
[b]Ref. [87]. Theoretical work using different supercells
[c]Ref. [88]. Theoretical work

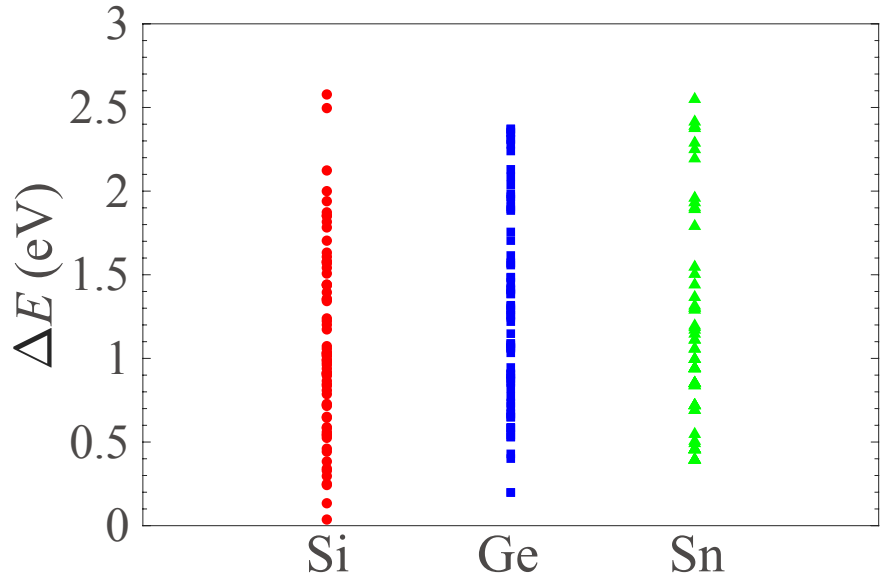

**Figure 6.** Formation energies $\Delta E$ of single Si, Ge, or Sn vacancies in the SiGeSn high-entropy alloy.

Figure 7(a) displays the dependence of vacancy formation energies on the number of nearest-neighbor Si and Ge atoms of a vacancy; the number of nearest Sn atoms equals four minus the former number, so it is not used as abscissa or ordinate in the plot. As can be seen from Figure 7(a), despite a vacancy has the same number of nearest-neighbor Si and Ge atoms (i.e., same abscissa and ordinate) the resulting formation energies can be quite different. For instance, 17 different energies are found if both abscissa and ordinate equal two. This large degeneracy in the number of nearest-neighbor Si atoms and Ge atoms implies that removing an atom to create a vacancy impacts more atoms than the nearest neighbors. Furthermore, bond reformation (see below) is not considered in this simple model. Figure 7(b)-(d) show the same dependence for Si, Ge, and Sn vacancies, respectively. For the Si case (Figure 7(a)), we observe that the highest vacancy formation energy occurs when the removed atom is surrounded by four Si atoms, which is consistent with what we shall see below that breaking a Si-Si bond requires the most energy in comparison with breaking a Si-Ge or a Si-Sn bond. We also observe that many low vacancy formation energies fall in the regions where the number of nearest-neighbor Si atoms is below two. For example, the lowest-vacancy-formation-energy structure corresponds to the one with a vacancy that has one nearest-neighbor Si atom, two Ge nearest-neighbor atoms, and one Sn nearest-neighbor atom, respectively. For the Ge case (Figure 7(b)), higher vacancy formation energies also tend to occur when the number of nearest-neighbor Si atoms is above or equal to two, while the lowest formation energy appears anomalously where there are three nearest-neighbor Si atoms in spite of the Si-Ge bond strength being the second strongest among the bonds formed from Si, Ge, and Sn atoms. The location of the highest vacancy formation energy for the Sn case (Figure 7(c)) is similar to that in the Ge case. In addition, the vacancies that have lower formation energies prefer to have fewer Si atoms as nearest neighbors, a common feature in all the three cases.

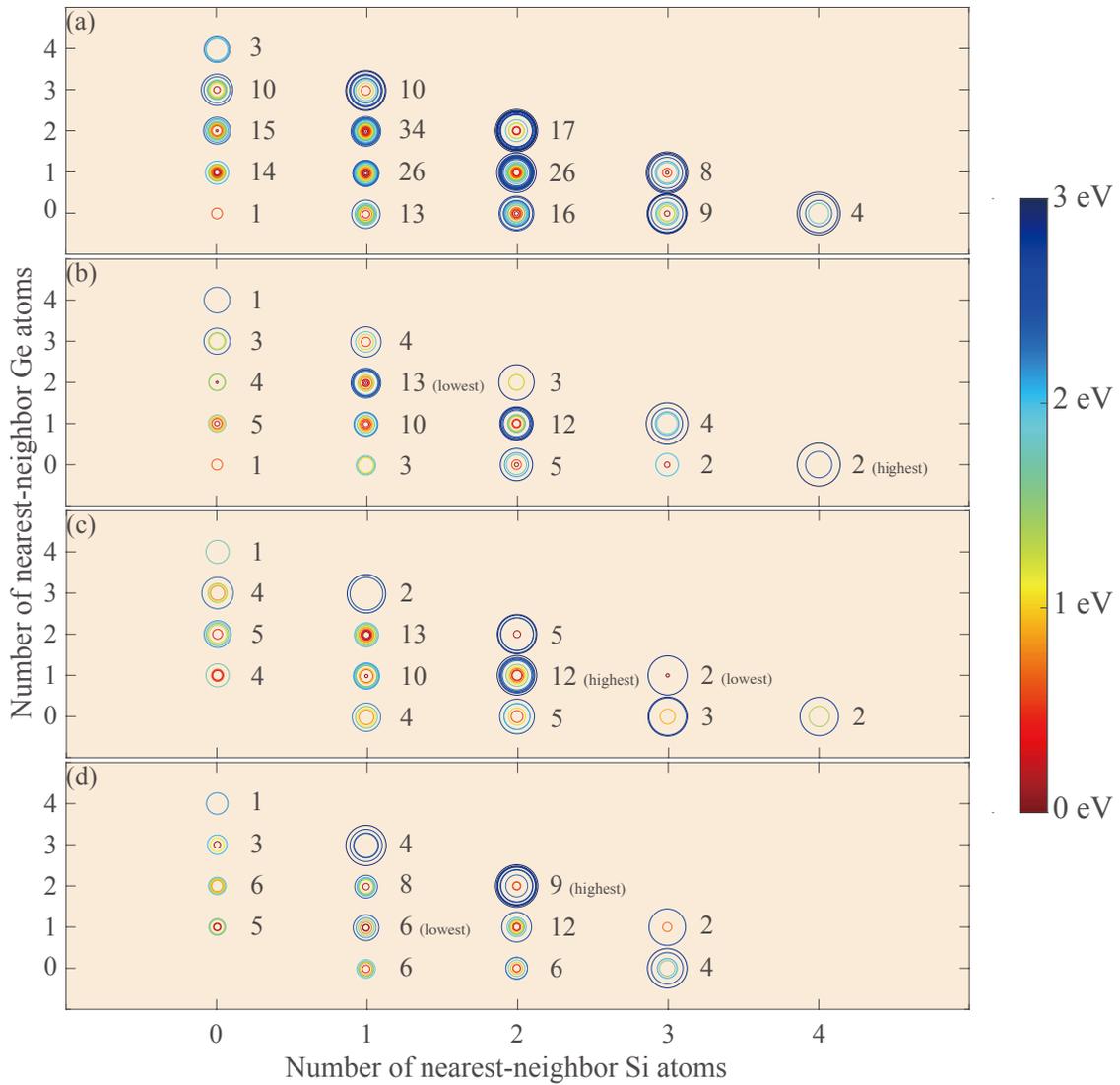

**Figure 7.** Relation between vacancy formation energies and the number of nearest-neighbor Si and Ge atoms near each single vacancy in the SiGeSn high-entropy alloy. (a) is for the entire data set and its division into the (b) Si (c) Ge, and (d) Sn subsets. The radius and color intensity of each circle represent the magnitude of the corresponding vacancy formation energy. The numbers near the circles denote the number of circles that share the same abscissa or ordinate. The texts "highest" and "lowest" label the highest and lowest vacancy formation energies, respectively.

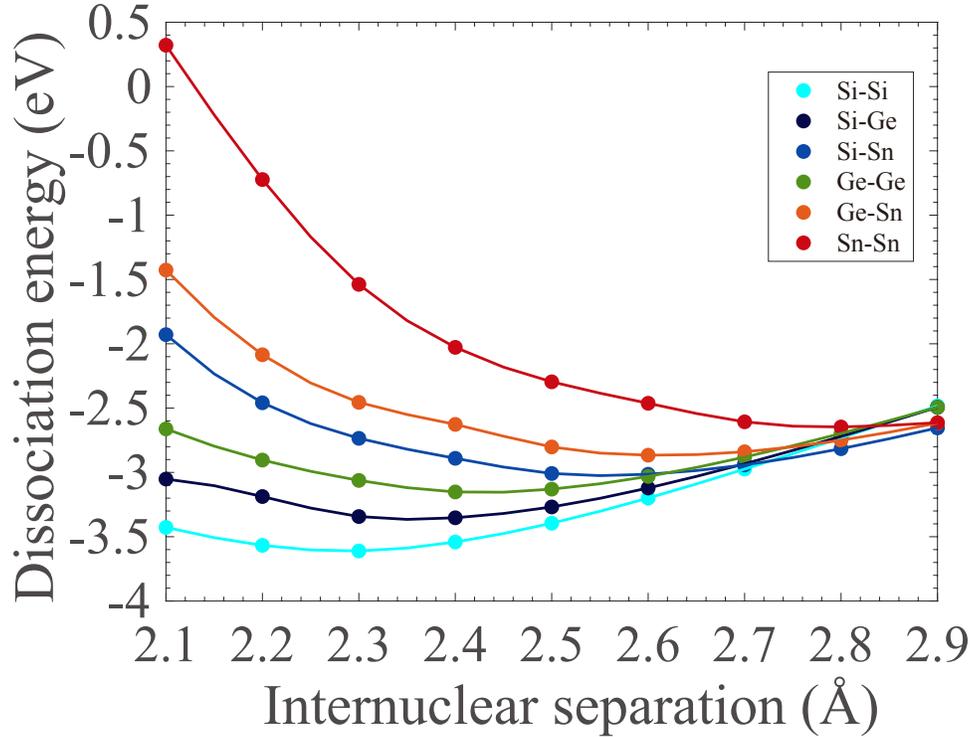

**Figure 8.** Dissociation energies of Si-Si, Si-Ge, Si-Sn, Ge-Ge, Ge-Sn, and Sn-Sn diatomics as a function of internuclear separation.

We also expect that the lower bounds of the three ranges of vacancy formation energies to be higher than the vacancy formation energy in bulk $\alpha$-Sn, i.e., 1.33 eV. This assumption is made again based on the fact that creating a vacancy is equivalent of breaking chemical bonds. The broken bonds in the SiGeSn HEAs are diversified, such as Si-Ge, Si-Sn, etc. The energies to break these mixed bonds are supposed to lie between the Si-Si and Sn-Sn bonds. To support our assumption, Figure 8 depicts the variation of dissociation energy with interatomic distance for Si-Si, Si-Si, Si-Ge, Si-Sn, Ge-Ge, Ge-Sn, and Sn-Sn diatomics. The dissociation energy is calculated as the energy difference of the total energy of a diatomic molecule (placed in the same vacuum box as used for computing the energies of isolated atoms) from the total energies of the isolated constituent atoms. The dissociation energies for Si-Si, Si-Ge, Si-Sn, Ge-Ge, Ge-Sn, and Sn-Sn are 3.61, 3.35, 3.01, 3.15, 2.87, and 2.65 eV, respectively, which are reasonably consistent with the experimental data of 3.31, 3.08, 2.42, 2.82, 2.38 and 1.99 eV [89-92], respectively; both theoretical and experimental data show that the Si-Si bond strength is the strongest while the Sn-Sn bond is the weakest. All the other bond strengths lie between these upper and lower bounds. The order of the dissociation energies is the same as that of the cohesive energies of bulk Si, Ge, and α-Sn.

Surprisingly, there are 2/3 (144 out of 216) of the vacancy formation energies in the whole data set that are smaller than the vacancy formation energy of bulk α-Sn (1.33 eV). To understand this, Figure 9 shows the local atomic structures (before and after geometry optimizations) near the vacancies with the highest and lowest vacancy formation energies for Si, Ge, and Sn. We observe a common phenomenon: for the structures with the highest formation energies, the number of dangling bonds (4) remains unchanged after geometry relaxations, similar to the situation where an atom is removed in bulk Si, Ge, and α-Sn. As a result, the energies of the structures as well as the vacancy formation energies are high. By contrast, in the structures with the lowest formation energies, the atoms with dangling bonds due to removing an atom form new bonds to minimize the number of dangling bonds, thereby reducing the system's energy. For example, Figure 9 (a) illustrates that a Ge atom relocates to the location of the Si vacancy reforming four-fold coordinated bonds with nearest-neighboring two Sn atoms, one Ge atom, and one Si atom. This bond reformation makes the corresponding vacancy formation energy negligibly small (0.04 eV). Similar bond reformations are also found near the Ge and Sn vacancies (see Figure 9 (a) and (b)), leading to the corresponding lowest vacancy formation energies. We speculate that similar bond reformation also exists in a vacancy of metallic HEAs. For example, the formation energy of a single Fe vacancy in bulk Fe is 1.58 eV (this energy is also the lower bound of the vacancy formation energies of bulk Fe, Cr, Co and Ni that form the FeCrCoNi HEA), while the lowest formation energy of a single Fe vacancy in the FeCrCoNi HEA is 0.72 eV. The ratio of these two energies (1.58/0.72 ≅ 2.19) is much larger than the corresponding ratio (1.33/0.04 ≅ 33.25) in the SiGeSn HEA, manifesting the unique structure and energetics of the vacanncies in semiconducting HEAs.

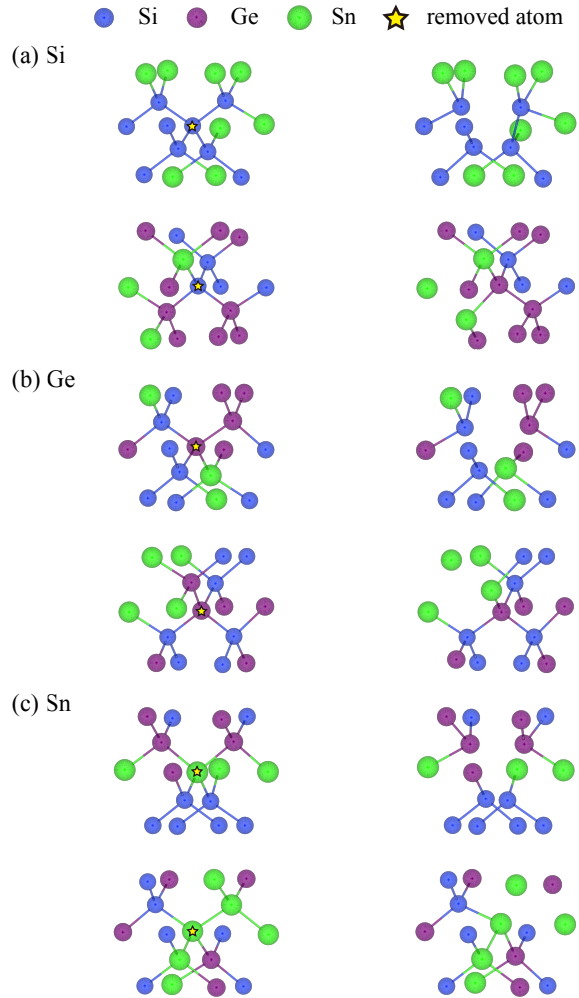

**Figure 9.** Local atomic structures near (a) Si, (b) Ge, and (c) Sn vacancies (denoted by yellow stars). The figures in the left and right columns of each panel show the atomic structures before and after geometry optimizations, respectively.

We now study the effects on the electronic structure of the SiGeSn HEA due to the presence of single vacancies. As shown above, it is necessary to use the HSE06 functional to obtain the more accurate electronic structures of the SiGeSn HEA. Because performing the calculations at the HSE06 level of theory for all the defected supercells—each has 215 atoms—is extremely time consuming, we compute the HSE06 electronic structures only for the six supercells with the lowest and highest vacancy formation energies for the three types (Si, Ge, and Sn) of vacancies, using their corresponding optimized configurations based on the PBE functional.

Figure 10 displays the density of states calculated for the vacancies with the highest and lowest formation energies computed with the HSE06 functional. A vacancy in a semiconductor usually leads to defect states in the band gap. For example, defect states resulted from a neutral oxygen vacancy are seen in the band gap of semiconducting $SrTiO_3$ [93]. We find no such defect states in the band gap of the SiGeSn HEA. Instead, we observe that all the lowest-formation-energy structures with single vacancies are semiconducting with band gaps of 0.31, 0.35, and 0.32 eV for the Si, Ge, and Sn vacancies, respectively, and that all the highest-formation-energy configurations are metallic. These band gaps are nearly identical to that (0.38 eV) of the SiGeSn HEA without a vacancy, implying trivial impacts of the vacancies on the band gap. The absence of the defect states in the lowest-formation-energy structures is because of the bond reformation as shown in Figure 9, which provides the otherwise missing orbitals to overlap with the orbitals from the atoms surrounding a vacancy site. Figure 9 shows that the vacancy sites in the lowest-formation-energy structures are replaced by another atom either of the different element (i.e., Si by Ge) or the same element (i.e., Ge by Ge and Sn by Sn), thereby the band gap size is almost not affected by these atomic replacements. In the highest-formation-energy vacancy supercells, because unreconstructed dangling bonds exist and their energies are large enough to excite the electrons to the energies above the valence band maximum. The excited electrons then form continuous bands bridging the band gap and resulting in overall metallic systems. These vacancies therefore render the SiGeSn HEA unsuitable for optoelectronic applications. Fortunately, we expect a low probability of having these vacancies as their formation energies are very high. In other words, the SiGeSn HEA will be dominated by those low-energy vacancies that retain the semiconducting properties required for doping purposes.

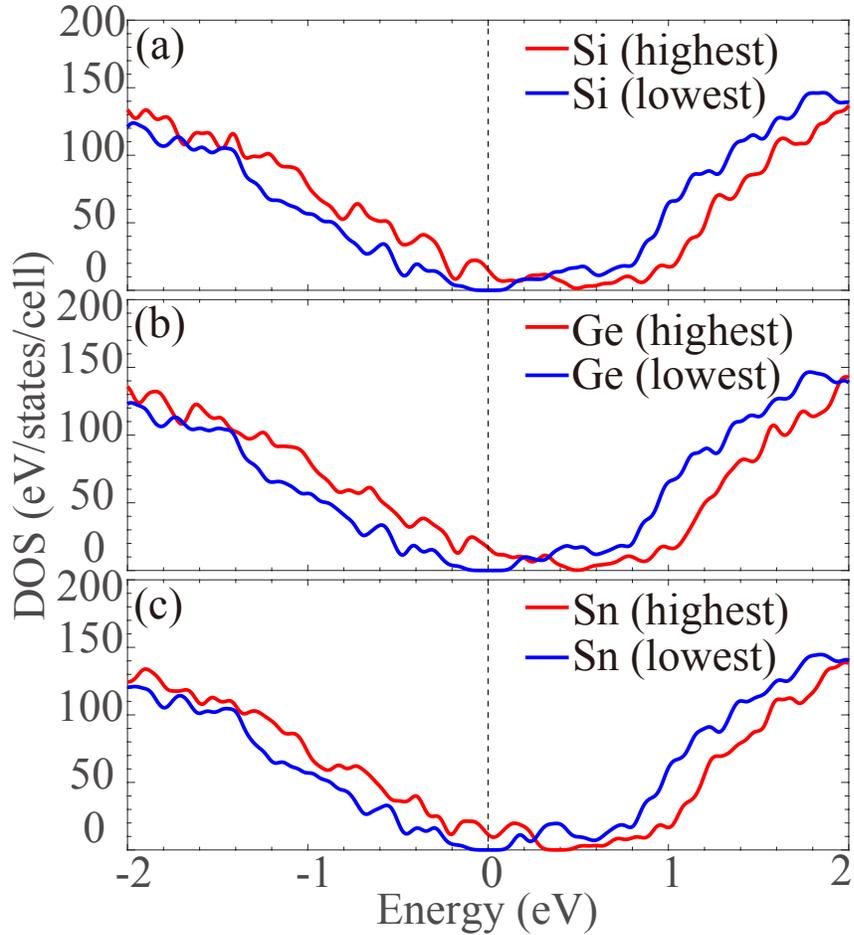

**Figure 10.** Density of states of (a) Si, (b) Ge, and (c) Sn of the SiGeSn high-entropy alloy with single vacancies of the highest and lowest formation energies.

**Conclusions**

We have studied a Si-Ge-Sn alloy in the context of HEA using DFT calculations. We showed that the SiGeSn HEA is a semiconductor with the band gap in the mid-infrared region. Comparing to metallic HEAs, the SiGeSn HEA exhibits the same, large lattice distortion effect and wide ranges of vacancy formation energies. Nevertheless, we also found two features that distinguish the SiGeSn HEA from metallic HEAs. First, the SiGeSn HEA has no CSRO due to the similar electronegativity of Si, Ge, and Sn, which consequently increases the configurational entropy.

Second, single vacancy formation energies of the SiGeSn HEA can be very low owing to the bond reformations near the vacancies. As a result, the band gap of the SiGeSn HEA is almost unaffected by the vacancies that have low formation energies. Overall, investigating Si-Ge-Sn alloys from the perspectives of HEAs will produce fundamental insight on the structure–property interplay of group IV alloys, and our integrated computational methods will assist the design of other groups of semiconducting HEAs exhibit critical electromagnetic properties and applications.

**Acknowledgements**

We thank the start-up funds from Arizona State University (ASU). W. H. thanks the Master's Opportunity for Research in Engineering (MORE) program as ASU. This research used computational resources of the Texas Advanced Computing Center under Contracts No.TG-DMR170070 and the Agave cluster at ASU.